\documentclass[%
reprint,
superscriptaddress,
 amsmath,amssymb,
prl,
longbibliography,
]{revtex4-2}

\usepackage{romannum}
\usepackage{graphicx}
\usepackage{dcolumn}
\usepackage{bm}

\usepackage{siunitx}

\usepackage{xcolor}
\definecolor{myred}{RGB}{216.7500,   82.8750,   24.9900}
\definecolor{myblue}{RGB}{0,  113.9850,  188.9550}
\definecolor{mydarkred}{RGB}{ 161.9250,   19.8900,   46.9200}
\definecolor{myyellow}{RGB}{236.8950,  176.9700,   31.8750}
\definecolor{mygreen}{RGB}{118.8300,  171.8700,   47.9400}

\usepackage{tikz}
\usetikzlibrary{shapes.geometric,shapes.symbols}

\begin{document}

\title{\large{Freezing-induced topological transition of double-emulsion}}

\author{Jochem G. Meijer}
\affiliation{Physics of Fluids group, Max Planck Center Twente for Complex Fluid Dynamics, Department of Science and Technology, Mesa+ Institute and J. M. Burgers Center for Fluid Dynamics, University of Twente, P.O. Box 217, 7500 AE Enschede, The Netherlands}

\author{Pallav Kant}%
\affiliation{School of Engineering, University of Manchester, M13 9PL, United Kingdom}

\author{Detlef Lohse}%
\affiliation{Physics of Fluids group, Max Planck Center Twente for Complex Fluid Dynamics, Department of Science and Technology, Mesa+ Institute and J. M. Burgers Center for Fluid Dynamics, University of Twente, P.O. Box 217, 7500 AE Enschede, The Netherlands}
\affiliation{Max Planck Institute for Dynamics and Self-Organization, Am Faßberg 17, 37077 Göttingen, Germany}

\begin{abstract}
Solidification of complex liquids is pertinent to numerous natural and industrial processes. 
Here, we examine the freezing of a W/O/W double-emulsion, i.e., water-in-oil compound droplets dispersed in water.
We show that the solidification of such hierarchical emulsions can trigger a topological transition; for example, in our case, we observe the transition from the stable W/O/W state to a (frozen) O/W single-emulsion configuration. 
Strikingly, this transition is characterised by sudden expulsion of the inner water drop from the encapsulating oil droplet. 
We propose that this topological transition is triggered by the freezing of the encapsulating oil droplet from the outside in, putting tension on the inner water drop thus, destabilizing the W/O/W configuration. 
Using high-speed imaging we characterize the destabilization process. 
Interestingly, we find that below a critical size of the inner drop, $R_{\mathrm{in,crit}} \approx \SI{19}{\micro \meter}$,  the topological transition does not occur any more and the double-emulsion remains stable, in line with our interpretation.
\end{abstract}

\date{\today}

\maketitle

The characteristic physical feature common between the fabrication of functionalized ceramics \cite{deville2006freezing,deville2007ice,deville2008freeze,deville2010freeze}, cryo-biology \cite{bronstein1981rejection,korber1988phenomena,muldrew2004water}, frost-heaving in colder regions \cite{rempel2010frost,peppin2011frost,peppin2013physics} and ice cream making is the interaction between insoluble dispersed particles in a melt and a moving solidification front.
Due to its relevance and impact in both natural phenomena and technological applications, this interaction has been extensively studied both experimentally and theoretically \cite{bronstein1981rejection,korber1988phenomena,shangguan1992analytical, rempel2001particle, muldrew2004water, park2006encapsulation, deville2006freezing, deville2007ice, rempel2010frost}.
The prominent focus of these investigations has been to determine the conditions under which dispersed particles are either engulfed or rejected by the advancing freezing front \cite{shangguan1992analytical,rempel2001particle,dedovets2018five, tyagi2020objects,meijer2023frozen}. 
This aspect holds particular importance in material science, as the distribution of dispersed particles profoundly influences the resulting micro-structure, thereby governing the functional properties of the solidified materials.
Importantly, these collective efforts have revealed that the interaction between a particle and the moving solidification front is influenced by distinct physical processes operating across different length scales. This characteristic renders it a true multi-scale problem, with each decade of length scale presenting its own set of governing mechanisms and phenomena.
For instance, at the particle level, the interaction is influenced by factors such as thermal conductivity mismatches between the particle and the melt, as well as density changes during the liquid-solid phase transition. At smaller scales, van der Waals interactions and the flow of lubricating films in premelted layers dictate the local dynamics \cite{wettlaufer2006premelting}. 
Furthermore, recent investigations involving "soft" particles, such as droplets or bubbles, have shown that the overall complexity is further amplified due to the mechanical deformation experienced by the particles during the interaction with the moving solidification front \cite{tyagi2021multiple,tyagi2022solute,meijer2023thin}.

The current understanding of this subject primarily stems from studying the interaction between \textit{uniform} particles and a moving solidification interface. However, in modern technological applications, such as cryobiology and material science, complex and \textit{multi-component} particles are frequently encountered. 
Inspired by these applications, in this Letter, we extend the envelope of the current state of the art on this subject to the interaction between the solidification front and \textit{multi-component} particles.
Using a dilute W/O/W double-emulsion as a model system, we report unexpected behaviors that emerge during its solidification.
We show that dispersed compound droplets in a double-emulsion can undergo a topological change upon engulfment in the solidifying bulk.
This topological transition is triggered by the partial solidification of the dispersed compound droplet.
Our findings are relevant for exerting better control over cryopreservation procedures of bio-specimen as well as the fabrication of advanced materials.

\begin{figure}[b!] 
\includegraphics[width=0.45\textwidth]{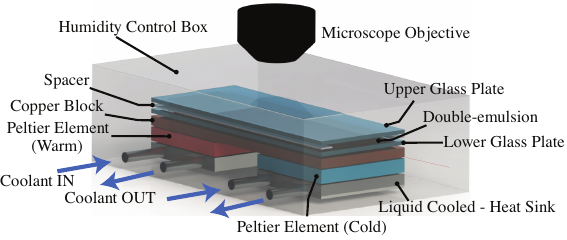}
\caption{ Schematic of the experimental setup. }
\label{fig:1}
\end{figure}

\begin{figure*}  
\includegraphics[width=0.85\textwidth]{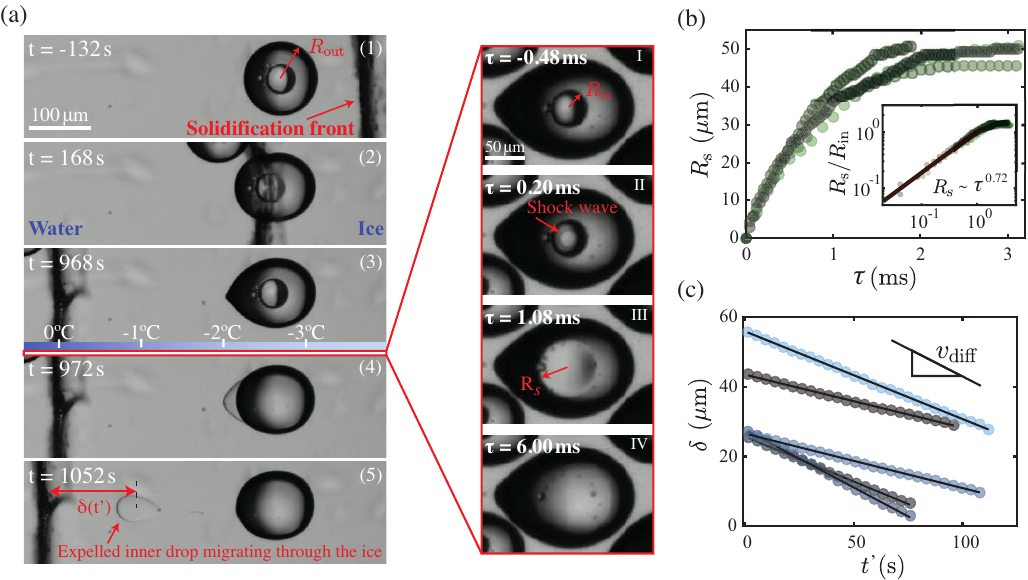}
\caption{(a) (1-5) Sequence of images capturing the engulfment of a water-in-oil compound droplet by a solidification front, advancing at rate $V = \SI{0.5}{\micro \meter \per \second}$ (Supplementary Movie 1).
The moment in time the front makes first contact with the droplet is set to $t=0$. 
The third panel shows a sketch of the expected temperature distribution in the ice at a certain moment in time set by the strength of the applied thermal gradient $G$.
(I-IV) High-speed images were acquired at 25.000 fps, showing the sudden topological transition (Supplementary Movies 2 \& 3). 
At $\tau = 0$ a shock wave appears at the center of the inner drop and an ethanol vapour cloud expands radially outward. 
(b) Radius $R_s$ of the shock wave as a function of time for different transitions, indicated by different shades.
The inset shows that at the early stages, when $R_s/R_{\mathrm{in}} < 1$, it follows a power-law-like behaviour that is consistent with that of a diverging spherical shock wave \cite{guderley1942,landau2013fluid}. 
(c) Distance $\delta$ between the solidification front and the center of the expelled inner drop migrating through the ice.
Note the $\sim 5$ orders of magnitude difference in time scale in (c) as compared to (b).   
At $t' = 0$ the expelled drop detaches from the oil. 
The migration velocity is determined by the velocity $v_{\mathrm{diff}}$ at which diffusion takes place within the expelled water-ethanol pocket.}
\label{fig:2}
\end{figure*}

The W/O/W double-emulsions used in our current experiments are prepared using a microfluidic setup similar to the one described in Ref.\,\cite{van2021feedback}.
In our adopted setup, we exploit the process of spontaneous liquid-liquid phase separation of a ternary mixture (water, ethanol and oil) to achieve double-emulsions, as discussed in detail in Refs.\,\cite{haase2014tailoring,moerman2018emulsion}.
Note that here we use diethyl-phthalate (DEP) as an oil phase and Pluronic-F127 as surfactant to stabilise the emulsion.
The oil has a freezing point close to that of water, i.e., $T_{\mathrm{m,DEP}} = \SI{-3}{\celsius}$.
Since the process of emulsion preparation is extremely sensitive to the initial composition of the ternary mixture, we use a DEP/ethanol/water mixture (0.74/0.05/0.21 vol.) with which we achieve a stable double-emulsion.
As a consequence of the adopted methodology of the double-emulsion preparation, both the inner water drop and the encapsulating oil droplet contain traces of ethanol.

The experimental assembly employed in current investigations is shown schematically in Fig.\,\ref{fig:1}.
In a typical experiment, the W/O/W double-emulsion is first injected into a horizontal Hele-Shaw cell of thickness $\SI{200}{\micro \meter}$, resting on a copper block.
A thermal gradient $G \sim \SI{5}{\kelvin \per \milli \meter}$ is then applied to this copper block using Peltier elements (spaced $\SI{3}{\mm}$ apart), such that the continuous phase of the double-emulsion (water) begins to solidify uniformly from the cold end.
Solidification is initiated by placing a small piece of ice at the cold end of the cell when its temperature falls below the melting point of water. 
It is important to note that, due to the extremely slow freezing rates employed in our experiments, clear ice is formed with a refractive index (1.31) similar to that of water (1.33).
The solidification rate $V$ of the bulk medium is controlled by the applied thermal gradient $G$.
The engulfment of the dispersed compound droplets into the solidified bulk is recorded in top-view using a camera (Nikon D850 or Photron NOVA S16) connected to a long working distance lens (Thorlabs, MVL12X12Z plus 2X lens attachment).
The sample is illuminated with diffused cold-LED to avoid localised heating. 
For further experimental details and the protocol for the double-emulsion preparation, we refer to the Supplemental Material.

The sequence of images in Fig.\,\ref{fig:2}\,(a) highlights the events leading up to the sudden topological transition of a W/O/W double-emulsion when a dispersed water-in-oil compound droplet is engulfed into the solidifying bulk at a steady rate, $V \approx \SI{0.5}{\micro \meter \per \second} $.
The moment in time when the solidification front makes contact with the compound droplet is defined as $t=0$. 
As shown in Fig.\ref{fig:2}\,(a), while the solidification front passes over, the encapsulating oil droplet severely deforms into a pointy, tear-like shape.
Recent findings \cite{meijer2023thin} have revealed that this deformation of the engulfed droplet is mediated by pressure variations in the nanometric thin liquid film, also referred to as premelted film \cite{wettlaufer2006premelting}, sandwiched between the dispersed droplet and the solidifying bulk.  
In contrast, during this engulfment process, the inner water drop clearly remains unaffected.
However, very unexpectedly, a few minutes ($\sim \SI{16}{\minute}$) after the dispersed compound droplet is completely engulfed in the solidified bulk (ice) and the solidification front is not in its vicinity, the inner water drop suddenly gets expelled outside the encapsulating oil droplet, altering the topology of the initially stable W/O/W double-emulsion.
The expelled liquid gathers in the thin premelted film enveloping the oil droplet.
Interestingly, the accumulation predominantly occurs at the warmer end, suggesting that the temperature gradient plays a role in this process.
More importantly, as this topological transition occurs, the pointy, tear-like shape of the oil droplet relaxes to its original unperturbed (spherical) shape.
Finally, we observe that the expelled liquid eventually begins to migrate towards the moving solidification front at a constant speed $v_\mathrm{diff} \sim 2\,\mu\mathrm{ms^{-1}}$, see Fig.\,\ref{fig:2}\,(c).
This surprising migration of the liquid through the solidified bulk resembles a process akin to brine-pocket diffusion in sea ice that is driven by a delicate diffusive process of solute diffusion and its influence on the local phase equilibrium \cite{whitman1926elimination, hoekstra1965migration, harrison1965measurement}.
It is worth noting that, as a result of the chosen fabrication method for the double-emulsions used in this study, the dispersed compound droplets contain traces of ethanol, which eventually acts as a solute, thus enabling a diffusive process that leads to the migration of the liquid pocket.
For a detailed discussion of the physical mechanisms leading to the migration of the liquid pocket, we refer to the Supplementary Material.
In the following, we focus on the main finding of this paper: the sudden topological transition of the double-emulsion.

Since the time scales associated with the topological change of engulfed compound droplets are much faster than the rate of solidification, we employ high-speed imaging to reveal the details of this apparently instantaneous event.
As illustrated in the sequence of images in Fig.\,\ref{fig:2}\,(a)(I-IV), the transition is triggered at the center of the inner drop, marked by the appearance of a radially expanding dark ring. 
Remarkably, this outward-moving dark ring is also accompanied by a localized increase in light intensity at its center, forming a discernible bright spot. 
Once the size of the bright spot exceeds the dimensions of the inner drop, it gradually fades (see Supplemental Material), coinciding with the complete disappearance of the inner water drop.
Following this event, the disappeared liquid collects outside the enveloping oil droplet within the premelted film in about $10-100\,\SI{}{\milli \second}$ (not shown here).
By tracking the temporal expansion of this dark ring, $R_s (t)$, we obtain a power-law-like behaviour $R_{\mathrm{s}} \sim \tau^{0.72}$ at the early stages of its growth when $R_s/R_{\mathrm{in}} < 1$, see inset Fig.\,\ref{fig:2}\,(b).
Such scaling behavior evokes similarities with an adiabatically expanding spherical shock wave, which entails a discontinuity in density, pressure, and temperature \cite{guderley1942,landau2013fluid}.
Considering that the double-emulsions used in our current experiments contain traces of ethanol, we propose that all the above described features are triggered by its phase transition from liquid to vapor during the solidification of the surrounding oil droplet.
Since ethanol is crucial for the stability of the double-emulsion, its sudden phase transition leads to the dissolution of the inner water drop, destabilizing the compound droplet.
The process of dissolution, we argue, is optically screened-off by the vapour cloud (bright spot) itself and therefore not visible in Fig.\,\ref{fig:2}\,(a)(I-IV).
Note that the ethanol vapor allows greater transmission of light because of its reduced refractive index, which in turn rationalizes the appearance of a bright spot during the transition.
Also, we will see later that the rate of dissolution indeed exceeds the rate at which the vapour cloud expands.

\begin{figure}[b]
\includegraphics[width=0.45\textwidth]{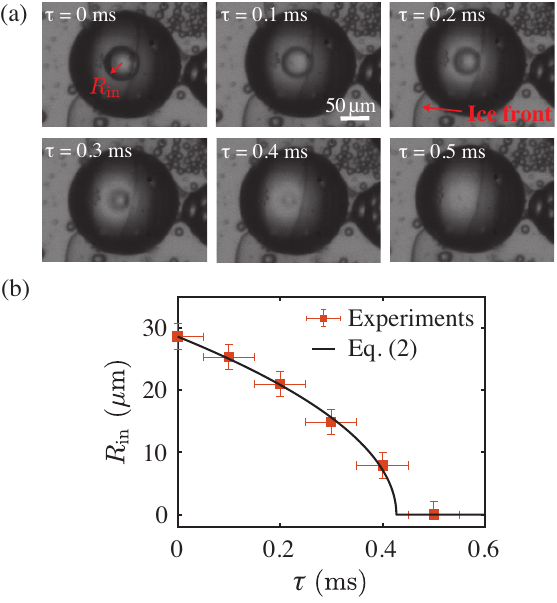}
\caption{(a) Experimental snapshots showing the sudden dissolution of the inner drop, here triggered by the ice front passing over the compound droplet (Supplementary Movie 4). (b) Radius $R_{\mathrm{in}}$ of the inner water drop as function of time (boxes). The rate of shrinking for several of such cases follows the steady Epstein-Plesset model of dissolution (solid line, Eq.\,(\ref{eq:EpsteinPlesset})). }
\label{fig:3}
\end{figure}

The hypothesized phase transition of ethanol during the solidification of the double-emulsion is rationalized as follows:
Following engulfment, as the solidification front steadily recedes away from the dispersed compound droplet, the local temperature in its vicinity progressively decreases in line with the applied thermal gradient (see sketch in Fig.\,\ref{fig:2}(a)(3)).
Upon reaching the freezing point of the oil ($T_{\mathrm{m,DEP}} = \SI{-3}{\celsius}$), the outer oil droplet begins to solidify, causing shrinkage in its volume, since $\rho_{\mathrm{oil-solid}}> \rho_{\mathrm{oil-melt}}$.
Nevertheless, this volumetric shrinkage is prevented due to the confinement by the solidified bulk water.
As a result the pressure inside the encapsulating oil droplet plummets, imposing tension on the inner water drop and causing it to stretch.
An approximate estimate of the pressure reduction experienced by the inner drop due to the solidification of the encapsulating oil droplet can be determined from the Gibbs-Duhem relation \cite{reif1965statistical}
\begin{equation}
\rho_{\mathrm{solid}} \,  \mathcal{L} \,  \frac{T_{\mathrm{m}} - T}{T_{\mathrm{m}}} = \Delta p \, \left(1 - \frac{\rho_{\mathrm{solid}}}{\rho_{\mathrm{melt}}} \right).
\label{Eq:Gibb-Duhem}
\end{equation}
Here,  $\mathcal{L}$ is the latent heat of solidification, $T$ is the temperature and $\Delta p$ the corresponding pressure change which we want to evaluate.
Estimating $\Delta p$ from Eq.\,(\ref{Eq:Gibb-Duhem}) with typical values, $T_{\mathrm{m}}-T = \SI{0.1}{\kelvin}$ and $1 - \rho_{\mathrm{solid}}/\rho_{\mathrm{liquid}} = \SI{-0.1}{}$, yields $\Delta p \sim \SI{-1}{\mega \pascal}$.
Such a depression in pressure critically affects the local phase equilibrium of ethanol for which the saturated vapour pressure at $T = \SI{-3}{\celsius}$ is $p_{v,\mathrm{eth}} \approx \SI{1.2}{\kilo \pascal}$.
To verify our hypothesis, we performed additional experiments by replacing DEP with dibutyl-phthalate (DBP) as an oil phase, while keeping the procedure of the double-emulsion preparation the same (see Supplementary Material). 
Since the freezing point of the DBP is much lower compared to DEP, $T_{\mathrm{m,DBP}} = \SI{-35}{\celsius}$, which cannot be achieved in our experimental setup, no abrupt topological transitions are observed. 
For further details related to these experiments, we refer to the Supplementary Material. 

Since the dissolution process of the inner water drop is screened-off by the ethanol vapor cloud (bright spot) appearing at the center, it is extremely difficult to directly measure related characteristic features.
For this, we rely on experiments performed with metastable W/O/W emulsions.
In these instances, especially for relatively large compound droplets with an inner drop size greater than $R_{\mathrm{in}}\geq \SI{30}{\micro \meter}$, the dissolution process is directly measured without any hindrance from the ethanol vapour cloud, see Fig.\,\ref{fig:3}\,(a).
Note that the sudden dissolution of the inner drop in these barely stable compound droplets is triggered by disturbances in the ambient conditions.
The sequence of images in Fig.\,\ref{fig:3}\,(a) exemplifies one such case, where the inner drop dissolves while the compound droplet is getting engulfed in the solidifying bulk.
The rate at which the inner drops dissolve, here directely measured, agrees well with the simplified Epstein-Plesset model \cite{epstein1950stability, duncan2006microdroplet},
\begin{equation}
\begin{aligned}
\frac{\mathrm{d}R_{\mathrm{in}}}{\mathrm{d}\tau} &= -\frac{\alpha}{\rho_w} \frac{1}{R_{\mathrm{in}}(\tau)}, \quad \text{thus} \quad R_{\mathrm{in}}^2 &= R_{0,\mathrm{in}}^2 - \frac{2 \alpha}{\rho_w} \tau,
\label{eq:EpsteinPlesset}
\end{aligned}
\end{equation}
with $\alpha = D \Delta c$, where $D$ and $\Delta c$ are the unknown diffusion coefficient and concentration gradient, respectively and $R_{0,\mathrm{in}}$ is the initial drop size.
Eq.\,(\ref{eq:EpsteinPlesset}) is fitted through the experimental data (see Fig.\,\ref{fig:3}\,(b)) to obtain $\alpha \approx \SI{9.6e-4}{\kg \per \meter \per \second}$.
A typical dissolution time scale arises as $\tau_d = \rho_w R_{\mathrm{in}}^2 / 2 \alpha \approx \SI{0.3}{\milli \second}$, which is significantly smaller compared to the time it takes for the ethanol vapour cloud to fully expand, see Fig.\,\ref{fig:2}\,(b).

\begin{figure}[t]
\includegraphics[width=0.35\textwidth]{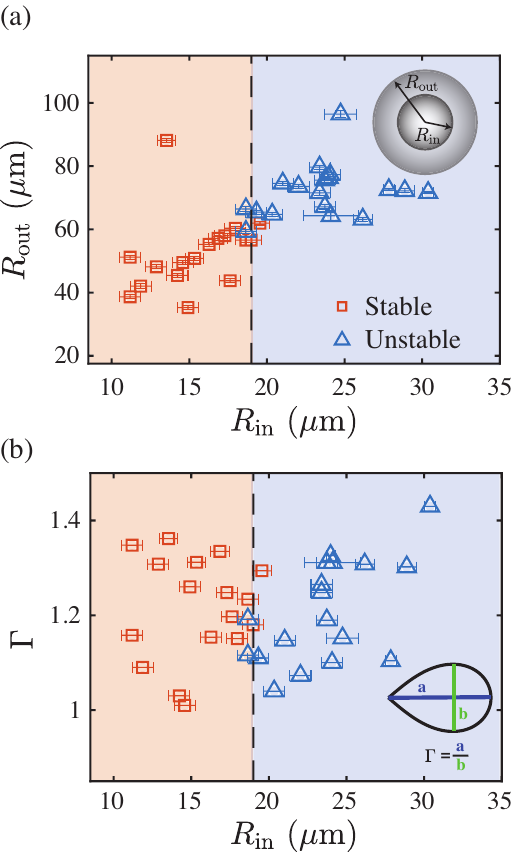}
\caption{Parameter space showing the dependence of the topological transition on (a) the size of the compound droplet and (b) its aspect ratio $\Gamma$ after encapsulation by the ice. The latter is directly related to the (adjustable) advancing velocity $V$ of the solidification front, i.e., $\Gamma \sim V^{-1}$ for pure oil drops in water \cite{tyagi2022solute,meijer2023thin}.  Below a critical inner drop size, $R_{\mathrm{in,crit}} \approx \SI{19}{\micro \meter}$, the compound droplets remain stable.
The rate of freezing does not affect this critical size.}
\label{fig:4}
\end{figure}

Beyond the freezing driven liquid-to-vapor phase transition of the constituent liquid, another surprising feature of our system is that the topological transition of the double-emulsions is strongly dependent on the size of the inner drop.
We find that below a critical size, $R_{\mathrm{in,crit}} \approx \SI{19}{\micro \meter}$,  the compound droplets remain stable, see Fig.\,\ref{fig:4}.
Note that due to the applied procedure of double-emulsion preparation $R_{\mathrm{out}}$ and $R_{\mathrm{in}}$ cannot be varied independently and their ratio, here $R_{\mathrm{out}}/R_{\mathrm{out}} \approx 3.3$ (see Fig.\,\ref{fig:4}\,(a)),  is governed by the ethanol mass transfer rate during preparation \cite{haase2014tailoring,moerman2018emulsion}. 
A second control parameter is the rate of freezing, characterised by the advancing velocity $V$. 
Due to experimental limitations we are not able to capture both the engulfment and sudden transition of each drop. 
To nevertheless still qualitatively address the rate at which freezing occurred, we couple the rate of engulfment $V$ to the experienced drop deformation $\Gamma$ once incorporated into the ice, as it has been established that both are inversely related \cite{tyagi2022solute, meijer2023thin}.
Here, a large deformation (larger $\Gamma$) is related to slower freezing (smaller $V$) and \textit{vice versa}, where for the range of advancing velocities is approximately $\SI{0.2}{\micro \meter \per \second} \lesssim V \lesssim \SI{2}{\micro \meter \per \second}$.
In Fig.\,\ref{fig:4}\,(b) we thus show that the critical size dependence remains unaffected by the extend of the experienced deformation $\Gamma$ and hence by the rate of engulfment $V$.

To examine this unexpected size dependence we consider that the pressure in the inner drop, $p_{\mathrm{in}}(t)$, needs to fall below the vapour pressure of ethanol, $p_{v,\mathrm{eth}}$, at $T = \SI{-3}{\celsius}$, to trigger the topological transition.
Due to the tension exerted by the partially solidifying oil on the inner drop a reduction in pressure is expected (ideally) as $\Delta p_{\mathrm{in}}(t) = - K \Delta V_{\mathrm{in}}(t)/ V_{\mathrm{in}}$, where $K$ is the elastic bulk modulus of water and $\Delta V_{\mathrm{in}}(t)/ V_{\mathrm{in}}$ the relative change in volume over time.
The initial pressure in the drop is given by the sum of ambient and Laplace pressure, resulting in the condition that the compound droplet remains stable if $p_{\mathrm{in,ref}} + 2 \sigma/R_{\mathrm{in}} - K \, \Delta V_{\mathrm{in}}(t)/V_{\mathrm{in}} > p_{v,\mathrm{eth}}$, or, in terms of the inner drop size, if $R_{\mathrm{in}} < 2 \sigma / \left( p_{v,\mathrm{eth}} + K \, \Delta V_{\mathrm{in}}(t)/V_{\mathrm{in}} - p_{\mathrm{in,ref}} \right)$, where $\sigma$ is the interfacial surface tension. 
Given the critical drop size, $R_{\mathrm{in,crit}}$, observed experimentally (see Fig.\,\ref{fig:4}) we conclude that the experienced change in volume over time, $\Delta V_{\mathrm{in}}(t)/ V_{\mathrm{in}}$, stagnates and that therefore the amount of tension applied to the inner drop is limited.
We suspect that, given the limitations of the experimental set-up, that sufficiently low temperatures could not be achieved, and the bounded applied thermal gradient, a thermal equilibrium is reached, causing only a small portion of the encapsulating oil to solidify. 

In summary, in this Letter, we discussed that a dilute double-emulsion, i.e., water-in-oil compound droplets in water, can undergo the topological transition from its W/O/W state to an O/W state after being frozen.
The sudden topological transition, we argue, is triggered by the partial solidification of the encapsulating oil, putting tension on the inner drop, causing the trace amounts of liquid ethanol inside to undergo a liquid-to-vapour phase transition.
As the vapour cloud expands,  ethanol is depleted, making the compound drop unstable, and the inner drop dissolves.
The then expelled interior accumulates at the warmer end and migrates though the ice in the direction of motion of the advancing solidification front.
The topological transition critically depends on the size of the inner drop and does not depend on the engulfment velocity $V$.
We found that below $R_{\mathrm{in,crit}} \approx \SI{19}{\micro \meter}$ the topological transition does not occur and the double-emulsion remains stable.
We rationalised that this size dependence originates from the limited amount of tension applied to the inner drop.
Our findings on the individual compound droplets can be extrapolated to those in a more densely packed double-emulsion, where the same phenomena is observed (see Supplementary Material).

The surprising topological transition we found is relevant when freezing \textit{multi-component} systems and in the context of the cryopreservation procedures of food emulsions and bio-specimen.
In addition, our findings highlight the effect of phase changing liquid inclusions in a solidifying material and how this might alter their topology in unexpected ways.

\section*{Acknowledgements}
The authors thank Gert-Wim Bruggert and Martin Bos for the technical support and Duco van Buuren for performing preliminary experiments. 
Additionally, we thank Andrea Prosperetti and Vincent Bertin for stimulating discussions.
The authors acknowledge the funding by Max Planck Center Twente and the Balzan Foundation.

\bibliography{References}

\clearpage
\renewcommand{\figurename}{Supplementary FIG.}
\setcounter{figure}{0}
\setcounter{equation}{0}
\renewcommand{\theequation}{S.\arabic{equation}}
\begin{widetext}
\section*{Supplementary Material}

\subsection*{List of Supplementary Movies}

\begin{itemize}
\item \textbf{Movie 1:} Movie showing the complete engulfment process of a compound droplet into the ice and the sudden topological transition, where the front advances at a velocity $V = \SI{0.5}{\micro \meter \per \second}$.  Snapshots are taken at a low frame rate of 0.25 frames/sec.  The movie corresponds to Fig.\,\ref{fig:2}(a)(1-5) of the main text. 
\item \textbf{Movie 2:} The sudden topological transition of a compound droplet completely surrounded by ice captured through high-speed imaging at 25.000 frames/sec. To ensure good optical resolution the field of view of the camera is limited and the moving solidification front can therefore not be captured simultaneously.  The movie corresponds to Fig.\,\ref{fig:2}(a)(I-IV) of the main text. 
\item \textbf{Movie 3:}  The sudden topological transition of a compound droplet completely surrounded by ice captured through high-speed imaging at 75.000 frames/sec. To ensure good optical resolution the field of view of the camera is limited and the moving solidification front can therefore not be captured simultaneously.  
\item \textbf{Movie 4:} The dissolution of the inner drop of a barely stable compound droplet ($R_{\mathrm{in}} > \SI{30}{\micro \meter}$) that is partially engulfed into the ice.  The process is captured through high-speed imaging at 10.000 frames/sec.  The movie corresponds to Fig.\,\ref{fig:3}(a) of the main text. 
\end{itemize}

\subsection*{General Experimental Details}

In this section we address some general experimental details and technicalities in more detail.
The preparation of the double-emulsion is discussed specifically in the following section.  
The exact protocol adopted in our current investigations is as follows. 
At the beginning of each experiment, the Hele-Shaw cell (Ibidi, $\mu$-Slide I Luer) is placed on a copper block maintained at a fixed thermal gradient and the double-emulsion is injected.
We then initiate the solidification on the colder end of the Hele-Shaw cell by inserting a small piece of ice, that acts as nucleation site and hence prevents the liquid to become supercooled.
After the initialization, a solidification front forms that slowly advances into the field of view of the camera. 
This can take easily up to an hour.
To capture the entire engulfment process we use a  Nikon D850 camera with long working distance lens (Thorlabs, MVL12X12Z) plus 2X lens attachment.
The working distance of the objective is $\SI{37}{\milli \meter}$ and 1 picture is taken every four seconds.
To visualise the sudden topological transition we use a high-speed camera (Photron NOVA S16) with the same objective and a frame-rate up to $\SI{75.000}{fps}$ (Supplementary Movie 3). 
The upper limit of the frame-rate is set by the amount of available back-light (Schott KL2500). \\
If the slide remained stationary, the front keeps advancing until it has propagated out of view. 
Alternatively, we can set the Hele-Shaw cell slide in motion with velocity V, opposite to the direction of motion of the solidification front, using a linear actuator (Physik Instrumente, M-230.25).
The lowest velocity that was measured at which the slide can be pushed is $V = \SI{0.1}{\micro \meter \per \second}$.  
The calibration measurements verified that at this velocity the variations are within 25\%. 
When increasing the velocity the stability increases and the variations fall within 10\% for $V = \SI{0.4}{\micro \meter \per \second}$ and beyond.
By properly choosing V it will appear as if the front remains stationary and that the droplets approach the front, whereas in reality it is vice versa. 
We therefore can change V by changing G.
To correct for irregularities in environmental heat losses, that slightly alter the growth rate of the ice, the voltage of the Peltier element on the colder side is continuously adjusted.
We do so to ensure that the solidification front remains stationary in our laboratory framework.     
Thus, although $V$ and $G$ can be altered independently, we chose to perform experiments in a manner that the velocity of the freezing front is determined by the applied thermal gradient.

\subsection*{Preparation of double-emulsion}

Here we discuss the preparation of the double-emulsion in more detail.
It is inspired by prior work \cite{haase2014tailoring,moerman2018emulsion}.
The double-emulsion is prepared using a 3D-printed microfluidic chip, schematically represented in Supplementary Fig.\,\ref{fig:SI_1}(a). 
Typically, for the preparation of single-emulsions, one uses a bulk phase (Milli-Q water) and a dispersed phase (oil).
Dependent on the flow rates of both phases through the microfluidic channels, a jet is formed that destabilises and thus produces sub-milimetric sized drops. 
Here, the oil is replaced by a ternary mixture of ethanol (Sigma-Aldrich, Germany), water (Milli-Q water) and diethyl phthalate (DEP) (Sigma-Aldrich, Germany), where the melting temperature of DEP is $T_{\mathrm{m,DEP}} = \SI{-3}{ \celsius}$.
Suitable flow rates are $Q = \SI{8}{\ml \per \minute}$ for the outer water phase and $Q = \SI{0.5}{\ml \per \minute}$ for the dispersed ternary mixture.

\begin{figure}[t!] 
\includegraphics[width=0.65\textwidth]{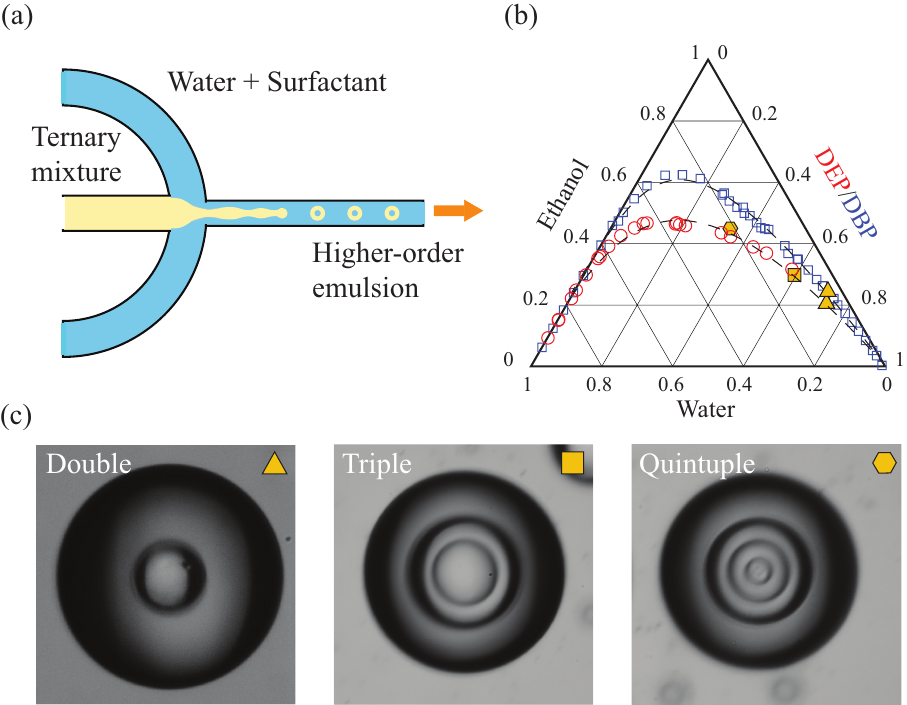}
\caption{(a) Schematic representation of the microfluidic device used to prepare the double-emulsion. the widths of the channels are approximately $\SI{0.7}{\mm}$. (b) Ternary phase diagram showing the binodal line of the ethanol/water/DEP (red) and ethanol/water/DBP (blue) systems \cite{bendova2001liquid}.  
Yellow symbols indicate the approximate initial composition of the ternary ethanol/water/DEP mixture to obtain a double-, triple or quintuple-compound droplet (c), respectively.
For our current study we use an initial composition of (0.21/0.05/0.74 vol.) for the ethanol/water/DEP system and (0.218/0.032/0.75 vol.) for the ethanol/water/DBP system.  }
\label{fig:SI_1}
\end{figure}

To ensure stability a surfactant Pluronic-F127 (Sigma-Aldrich, Germany) is added to the water.
In the initial phase of our study samples were prepared using different concentrations and dilutions to study the effect the surfactant concentration might have on the observations.  
For the dilutions that were considered we have observed that too few surfactant ($0.01\, \mathrm{wt}\%$) significantly reduces the lifetime of the compound droplets in the double-emulsion and too much surfactant ($1\, \mathrm{wt}\%$) causes the freezing front to become unstable.
The latter is also greatly affected by the amount of ethanol in the system. 
For our current study we have chosen an initial surfactant concentration of $0.1\, \mathrm{wt}\%$.
At the beginning of the experiment, before the sample is injected into the Hele-Shaw cell, it is diluted ten times with water (Mili-Q water) to ensure that the traces of ethanol in the system do not affect the stability of the growing crystal. 
The diluted double-emulsion sample remained stable for several days. \\
Dependent on the initial concentration of the ternary liquid, the drops that pinch off from the jet stabilise into higher order emulsions through liquid-liquid phase separation \cite{haase2014tailoring}.
Supplementary Fig.\,\ref{fig:SI_1}(b) shows the ternary diagram of the ethanol/water/DEP system \cite{bendova2001liquid}.
The different symbols indicate the approximate initial concentration of the ternary liquid and the double-, triple- and quintuple-emulsions that are produced; see Supplementary Fig.\,\ref{fig:SI_1}(c).
For our current study we have chosen an initial condition of (0.21/0.05/0.74 vol.) for ethanol, water and DEP, respectively, and hence focus on double-emulsions only.

\subsection*{Liquid pocket migration}

As discussed in the main text, after the topological transition has taken place, the expelled liquid accumulates at the warmer end of the engulfed droplet, before migrating at a constant migration velocity $\vert v_{\mathrm{diff}} \vert \approx 1-3\, \SI{}{\micro \meter \per \second}$ through the solidified bulk in the direction of motion of the solidification front, see Supplementary Fig.\,\ref{fig:SI_2}.
Recalling that the expelled water drop contains traces of ethanol and given the applied temperature gradient,
we argue that the migration process is equivalent to the process of brine pockets migrating through sea ice, i.e., 'brine pocket diffusion' \cite{whitman1926elimination, hoekstra1965migration, harrison1965measurement}.
The driving force for such migration originates from the gradient in composition over the pocket, which is governed by the applied thermal gradient $\mathrm{d}T/\mathrm{d}x$ and concentration dependence on temperature, $\mathrm{d}c/\mathrm{d}T$, given by its phase diagram.  
Under steady-state conditions, the flux of ethanol $c$ within the pocket is driven by diffusion. 
Due to this diffusion, the local compositions vary, as do the local, composition-dependent melting temperatures.
This causes water to freeze at the colder side and ice to melt at the warmer side. 
The velocity at which the water-ethanol pocket migrates, $v_{\mathrm{diff}}$, is thus given by a simple diffusion equation, $c \, v_{\mathrm{diff}} = -D \, \mathrm{d}c/\mathrm{d}x $, where $\mathrm{d}c/\mathrm{d}x = (\mathrm{d}c/\mathrm{d}T)(\mathrm{d}T/\mathrm{d}x)$.
Hence,
\begin{equation}
\left \vert \frac{\mathrm{d} \delta}{\mathrm{d} t'} \right \vert= \left \vert  v_{\mathrm{diff}} \right \vert = - \frac{D}{c}\frac{\mathrm{d}c}{\mathrm{d}T}G,
\end{equation}
where $D$ is the solute diffusivity, $\mathrm{d}T/\mathrm{d}x = G$ the applied thermal gradient and $c$ the ethanol concentration within the migrating volume.  
Considering that $D \sim \SI{1e-9}{\meter \squared \per \second}$, $c \sim \SI{0.02}{}$, $\mathrm{d}c/\mathrm{d}T \sim \SI{-0.02}{\per \kelvin}$ and $G \sim \SI{5}{\kelvin \per \milli \meter}$ yields $\left \vert  v_{\mathrm{diff}} \right \vert  \sim \SI{5}{\micro \meter \per \second}$, which agrees with the values observed experimentally.

\begin{figure}[t] 
\includegraphics[width=0.5\textwidth]{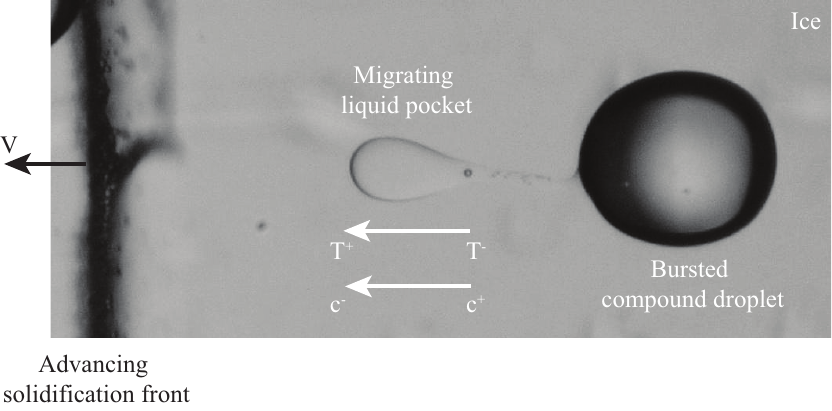}
\caption{Liquid pocket migration of the expelled interior through the ice. The applied thermal gradient induces a concentration gradient. While diffusion restores this concentration gradient, the local phase equilibrium at the colder and warmer sides are altered leading to the migration. Eventually, the pocket catches up with the ice front, which propagates with velocity $V$.}
\label{fig:SI_2}
\end{figure}

\subsection*{Light intensity increase during bursting}

The captured high-speed images reveal that during the radial expansion of the spherical dark feature a local increase in light intensity is observed.
The behaviour of the mean light intensity during one bursting event is measured and shown in Supplementary Fig.\,\ref{fig:SI_3}. 
When the dark feature first appears in the center of the inner drop, the intensity slightly decreases (I).
Then the mean intensity rapidly increases together with the expanding ring until its size coincides with the size of the inner drop (II), where a second minor decrease in intensity is measured (III).
Finally, the intensity reaches a certain maximum (IV) before slowly decaying towards an equilibrium value.

The origin of this overshoot in light intensity, we argue, is the change in optical path of the back-light as the ethanol vapour cloud expands, allowing more light to pass through to the camera without being deflected.
The interior of the drop thus appears brighter due to the lower index of reflection of the vapour.

\begin{figure}[t] 
\includegraphics[width=0.5\textwidth]{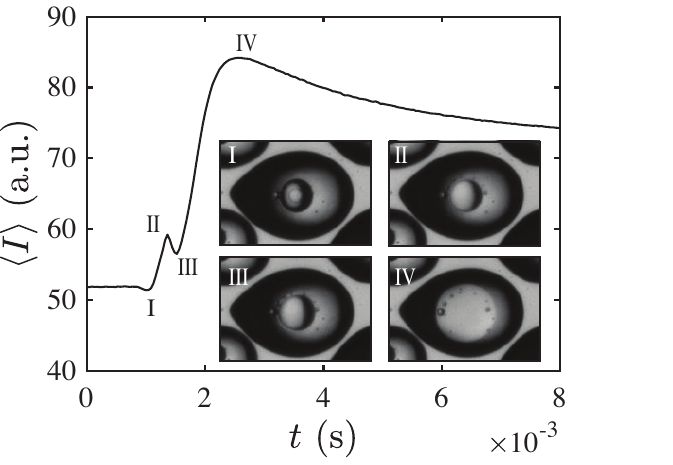}
\caption{ Mean of the light intensity captured during the topological transition. The inset shows the corresponding experimental snapshots. }
\label{fig:SI_3}
\end{figure}

\subsection*{Repeating experiments with dibutyl phthalate (DBP) }

To verify whether the partial solidification of the oil phase indeed triggers the topological transition, the experiments are repeated by replacing DEP with dibutyl-phthalate (DBP), that has similar properties compared to DEP but a significantly lower melting temperature, i.e.,  $T_{\mathrm{m,DBP}} = \SI{-35}{ \celsius}$.
These temperatures can not be achieved in our experimental setup.
The phase diagram of the ethanol/water/DBP system is shown in Supplementary Fig.\,\ref{fig:1} \cite{bendova2001liquid}.
The initial concentration is (0.218/0.032/0.75 vol.) for ethanol, water and DBP, respectively.
During the performed experiments using this system, no sudden topological transition of the double-emulsion was observed.

\subsection*{Freezing of a more densely packed double-emulsion }

Our current study focusses on the freezing of a dilute double-emulsion, meaning that the dispersed compound droplets are virtually isolated to study the behaviour of individual droplets that are being approached by an advancing solidification front. 
When considering the freezing of a more densely packed double-emulsion, where compound droplets are closely together (see Supplementary Fig.\,\ref{fig:SI_4}), we observe the same phenomena as discussed in the main text for those compound drops where $R_{\mathrm{in}} > R_{\mathrm{in,crit}}$, once they are incorporated in the ice and the front has advanced far enough (see red rectangles in Supplementary Fig.\,\ref{fig:SI_4}). 

\begin{figure}[t!] 
\includegraphics[width=0.65\textwidth]{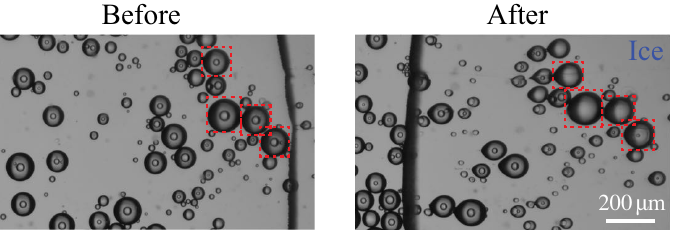}
\caption{Experimental snapshots of before and after the freezing of a more densely packed double-emulsion. The compound droplets incorporated in the ice for which $R_{\mathrm{in}} > R_{\mathrm{in,crit}}$ will eventually expel their inner core (see red rectangles).}
\label{fig:SI_4}
\end{figure}

\end{widetext}

\end{document}